# Multiple Location Profiling for Users and Relationships from Social Network and Content[*]

Rui Li[†]
ruili1@illinois.edu

Shengjie Wang[†]
wang260@illinois.edu

Kevin Chen-Chuan Chang[†,*]
kcchang@illinois.edu

[†] Department of Computer Science, University of Illinois at Urbana-Champaign, Urbana, IL, USA
[*] Advanced Digital Sciences Center, Illinois at Singapore, Singapore

## ABSTRACT

Users' locations are important for many applications such as personalized search and localized content delivery. In this paper, we study the problem of profiling Twitter users' locations with their following network and tweets. We propose a multiple location profiling model (*MLP*), which has three key features: 1) it formally models how likely a user follows another user given their locations and how likely a user tweets a venue given his location, 2) it fundamentally captures that a user has multiple locations and his following relationships and tweeted venues can be related to any of his locations, and some of them are even noisy, and 3) it novelly utilizes the home locations of some users as partial supervision. As a result, *MLP* not only discovers users' locations *accurately* and *completely*, but also "explains" each following relationship by revealing users' true locations in the relationship. Experiments on a large-scale data set demonstrate those advantages. Particularly, 1) for predicting users' home locations, *MLP* successfully places 62% users and outperforms two state-of-the-art methods by 10% in accuracy, 2) for discovering users' multiple locations, *MLP* improves the baseline methods by 14% in recall, and 3) for explaining following relationships, *MLP* achieves 57% accuracy.

## 1. INTRODUCTION

Users' locations are important information for many advanced information services, such as delivering localized news, recommending friends and serving targeted ads.

Recently, social network sites, such as Facebook and Twitter, become important platforms for users to connect with friends and share information. For example, Twitter, a social network for users to follow others and publish tweets, now has 140 million active users and generates 340 million tweets daily. However, for most of users on these sites, their locations are missing. For example, on Twitter, only a few users (16%) register city level locations (*e.g.*, Los Angeles, CA). Most of them leave nonsensical (*e.g.*, "my home"), general (*e.g.*, "CA") or even blank information. Although Twitter supports GPS tags in tweets, even fewer users (0.5%) use this feature due to obvious privacy concerns.

[*]This material is based upon work partially supported by NSF Grant IIS 1018723, the Advanced Digital Science Center of the University of Illinois at Urbana-Champaign and the Multimodal Information Access and Synthesis Center at UIUC. Any opinions, findings, and conclusions or recommendations expressed in this publication are those of the author(s) and do not necessarily reflect the views of the funding agencies.

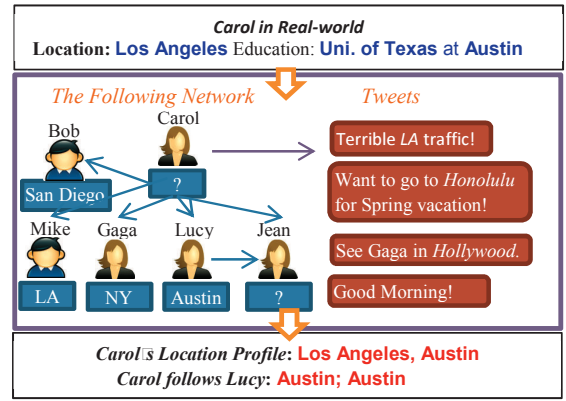

**Figure 1: Building Location Profiles for Users**

In the literature, many methods [8, 5, 11] have been proposed to profile users' locations in the context of social network. Specifically, they focus on profiling a user's *home location*, which is the single "permanent" resident location of the user, by exploring her social network (*e.g.*, friendships) and content (*e.g.*, tweets). Intuitively, both types of data provide valuable signals for profiling users' locations, as a user is likely to 1) connect to others living close to her, and 2) tweet her nearby "venues".

However, these methods have the same shortcoming – they assume that a user has only a "home location". In reality, as illustrated in Fig. 1, a user (*e.g.*, Carol) is related to multiple locations, such as her home location (*e.g.*, Los Angeles) and college location (*e.g.*, Austin). She follows friends from and tweets venues about all of them. *E.g.*, Carol follows her classmate Lucy in Austin and her co-worker Bob in Los Angeles. Thus, these methods not only profile her locations *incompletely*, but also estimate her home location *inaccurately*, because signals related to her other locations are noises for profiling even just her home location.

In this paper, we aim to build complete "location profiles" for Twitter users with their following network and tweets. We define a user's (*e.g.*, Carol) *location profile* as a set of locations related to her (*e.g.*, {Los Angeles, Austin}). It includes not only her home location (*e.g.*, Los Angeles) but also her other related locations (*e.g.*, Austin). Further, we clarify that each user related location is 1) a geo *scope* (*e.g.*, Los Angeles) instead of a geo *point* (*e.g.*, the Starbucks on 5th Ave.), and 2) a *long-term* location instead of a *temporally* related location (*e.g.*, the places where he is traveling). Thus, a user's location profile captures her multiple long-term geographic scopes of interests. We emphasize that we only use users' following network and tweets, and do not use GPS tags because they are rarely available as we just mentioned. Thus, we avoid the need for private information (*e.g.*, IP address) and enable third-party services (*e.g.*, researchers) to profile users' locations with Twitter open APIs.

In addition, for each relationship (*e.g.*, the following relationship from Carol to Lucy), we aim to profile users' specific locations underlying the relationship (*e.g.*, Carol follows Lucy as they studied in Austin), because a user has multiple locations of interest and each of her relationships can be a result of any of her locations. Profiling locations for each relationship not only helps us to discover users' locations accurately and completely, but also enables interesting applications, such as understanding the true geo connection between two users and grouping a user's friends into geo groups (*e.g.*, Carol is in Lucy's Austin group).

Thus, we propose a multiple location profiling model (*MLP*) for users and their relationships. To the best of our knowledge, *MLP* is the first model that 1) discovers users' multiple locations and 2) profiles both users and their relationships.

Specifically, *MLP* takes a generative probabilistic approach and models the *joint probability* of generating "following" and "tweeting" relationships based on users' multiple locations. With the joint probability, we estimate users' locations and locations of relationships as latent variables in the probability. However, when modeling the joint probability, *MLP* must deal with the following challenges.

**Location-based Generation** To connect users' locations with observed relationships, *MLP* needs to formally model the probability that a relationship is generated based on users' locations. Specifically, it should capture that a user at a specific location 1) follows her friends from different locations or tweets different venues, and 2) is likely to follow users living close to her or tweet her nearby venues.

We thoroughly investigate the connections between the two types of relationships and users' locations on a large-scale Twitter data and derive a *location-based generative model* for each type of relationships. For the "following probability" based on two user's locations, we explore the probability based on their distance, and formally model the probabilities over distances as a *power law distribution*. For the "tweeting probability" based on one user's location, we view locations and venues as discrete labels, and formally model the probabilities of tweeting different venues at each location as a *multinomial distribution* over a set of venues.

**Mixture of Observations** We can not straightforwardly use observed relationships to build a user' location profile, because of two challenges: 1) *the noisy-signal challenge*, which means she may follow friends (*e.g.*, Lady Gaga) and tweet venues (*e.g.*, Honolulu) that are not based on her locations, 2) *the mixed-signal challenge*, which means she follows friends (*e.g.*, Lucy and Bob) or tweets venues based on her multiple locations. We introduce two mixtures in *MLP* to deal with the two challenges.

With respect to the noisy-signal challenge, we model relationships as a mixture of "noisy" and "location-based" relationships. Specifically, we introduce a *random generative model* to model how a noisy relationship is generated randomly, besides the location-based generative model introduced above. Each relationship is generated by either of the two models with a certain probability. Thus, *MLP* explicitly captures noisy relationships, and automatically rules out them when profiling users' locations.

With respect to the mixed-signal challenge, we extend the location-based generative models to generate relationships based on users' multiple locations. Specifically, we view a user's *location profile* as a multinomial distribution over a set of locations, and extend the models to generate a location-based relationship in two steps: 1) generate a location assignment from each related user's location profile, and 2) generate the relationship based on the assignments. Thus, *MLP* fundamentally captures that a user has multiple locations. It not only discovers her multiple locations completely, but also estimates her home location accurately. Further, *MLP* reveals the true geo connection in a relationship with the location assignments for the relationship.

**Partially Available Supervision** As we mentioned that some users provide their home locations, those locations are the only observed locations and crucial for accurate profiling. However, they are difficult to use, because we can neither view them as users' location profiles, as a profile should contain more than a home location, nor use them to generate relationships because of the mixed-signal challenge.

We incorporate the observed home locations as prior knowledge to generate users' location profiles. Specifically, we assume that a user's location profile is generated via a prior distribution with a hyper parameter, and use the observed locations to set the hyper parameter for each user. As a result, for a user with an observed location, her derived location profile has a large probability to generate the observed location, and her relationships are likely to be generated based on the location as well.

Based on *MLP*, we profile users and their relationships as estimating the latent variables in the joint probability. However, as *MLP* models the above new aspects and integrates discrete (multinomial) and continuous (power low) distributions, it does not allow exact inference. We derive an efficient sampling-based algorithm based on the Gibbs sampling framework to estimate the latent variables.

To evaluate *MLP*, we conduct extensive experiments and compare *MLP* with the stare-of-the-art methods [5, 8] on a large-scale Twitter data containing about 160K users. The results show that *MLP* is effective. Specifically, 1) for predicting users' home locations, *MLP* largely improves the baseline methods by 10% and places 62% users accurately; 2) for discovering users' multiple locations, *MLP* captures users' multiple locations accurately and completely, and improves the baseline methods by 11% and 14% in terms of "precision" and "recall"; 3) for explaining following relationships, *MLP* achieves 57% accuracy.

The rest of this paper is organized as follows. We review the literature in Sec. 2, formalize our problem in Sec. 3, and develop our model in Sec. 4. Finally, we present experiments in Sec. 5 and conclude our work in Sec. 6.

## 2. RELATED WORK

In this section, we discuss some related work. In terms of the problem, our work is related to location prediction. In terms of the technique, our work is related to collective classification and mixture models.

**Location Prediction** As we focus on profiling users' locations, our work is related to identifying the geographical scope of various kinds of online resources, such as pages [10, 2], queries [4], tags [17], and photos [9]. However, they predict locations for different types of entities with different resources. For example, Amitay et al. [2] explore a web page's content to predict its geo scope via heuristically associating extracted location signals (*e.g.*, city names) to locations with a gazetteer. Our work is different, as we take a probabilistic approach to profile users' locations. Backstrom et al. [4] use a probabilistic model to assign a geographic center to a query based on its usage. Our probabilistic model is different, as it models generating following relationships and tweeted venues based on users' locations and assumes that a user has multiple locations.

Our work is most related to [8, 5, 11], as they profile users' locations as well. Cheng et al. [8] estimate a user's location based on his tweets. They identify a set of location related words (*e.g.*, "houston") and use these words as features to classify a user to locations. Backstrom et al. [5] estimate a user's location based on his friends on Facebook. They learn a function which assigns the probability of being friends given the distance of two users, and then estimate a user's location based on the maximum likelihood estimation principle. Recently, we propose a generative model to integrate both social network and tweets [11]. However, as we discussed in Sec. 1, as those methods assume a user has only one location, they not only profile a user's locations *incompletely*, but also estimate his home location *inaccurately*.

**Collective Classification** As we aim to assign users in a social network to location labels, our work is related to collective classification [18], which classifies objects in a network setting. For example, in [13], the authors take a local consistent assumption that a node's label is likely to be the same as its neighbors, and derive a voting-based neighborhood classifier. In [20], the authors apply a Markov dependency assumption that the label of one node depends on its neighbors' labels, and develop a pairwise Markov random field model. However, those methods will fail in our setting because of two reasons. First, they fail to utilize distances between location labels to make accurate classification. *E.g.*, given a user, who has three friends in New York, Los Angeles and Santa Monica respectively, a voting-based classifier assigns the user to the three locations with the same probability. if we capture that Los Angeles and Santa Monica are close, we are able to assign the user to Los Angeles area. Second, they assume that 1) a node has one label, and 2) all of its relationships are related to the label. Thus, they fail to address the mixed-singal challenge and will profile users' locations inaccurately and incompletely.

**Mixture Models** In terms of modeling observations (*i.e.*, relationships and tweeted venues) as generated by a mixture of hidden variables (*i.e.*, locations), *MLP* works in a similar way as Latent Dirichlet Allocation (LDA) [7] and Mixed Membership Stochastic Blockmodels (MMSB) [1].

LDA and its various extensions [19, 21] model a text collection as a mixture over a set of hidden topics. There are clear distinctions between *MLP* and LDA. First, *MLP* models *locations* instead of *topics* as the variables. Locations are predefined attributes, which can be observed from some users and have explicit correlation, while topics are loosely defined "clusters" of tokens, which are hidden in documents. In order to *classify* users into location labels, *MLP* explores distances between locations and utilizes observed locations from some users as supervision. Second, *MLP* models following relationships in addition to content (tweeted venues), as observations. We introduce a new generative process and a new probabilistic distribution (power law) to model them.

MMSB and its extensions [15] explicitly model how relationships (*e.g.*, citations) are generated based on a mixture of nodes' communities (*e.g.*, papers' topics). As communities are also loosely defined clusters, *MLP* is different from it by the first reason mentioned above. Furthermore, *MLP* advances MMSB in modeling relationships as well. MMSB assumes that a relationship between two nodes is generated based on pairwise interactions of their communities, while *MLP* explicitly explores the correlations between locations and introduces a power law distribution over distances to parameterize pairwise location interactions. As a result, we greatly eliminate the number of parameters and explicitly capture that users in a following relationship are likely to live close (see details in Sec. 4.4).

## 3. PROBLEM ABSTRACTION

In this section, we first introduce Twitter, and then abstract our problem from there.

As illustrated by Fig 1, Twitter is a social network, where users follow others and tweet messages. Typically, a user $u_i$ (*e.g.*, Carol) in Twitter connects to two types of resources, 1) her *following network*, which is a set of users (*e.g.*, Bob and Lucy), who follow or are followed by the user, and 2) her *tweeting content*, which is a set of messages tweeted by the user. Every $u_i$ is related to a set of locations, which is $u_i$'s location profile, denoted as $L_{u_i}$. $L_{u_i}$ contains $u_i$'s home location (*e.g.*, Bob's home location San Diego), denoted as $l_{u_i}$, and other related locations. Our goal is to build the location profile for each user, and we are interested in profiling their city-level locations specifically. All possible city-level locations can be given by a gazetteer, which can be easily obtained from various online resources (*e.g.*, Geographic Names Information System). We name them as candidate locations, and use $L$ to denote them. Further, some users' home locations are observed. We call them as labeled users, denoted as $U^*$, and the remaining users as unlabeled users, denoted as $U^N$. We use $U$ to denote all the users, where $U = U^* \cup U^N$.

As mentioned in Sec. 1, both types of resources are useful for profiling a user's locations, because a user (*e.g.*, Carol) is likely to 1) follow and be followed by users (*e.g.*, Mike and Bob), who live close to her, and 2) tweet some "venue names" (*e.g.*, Los Angeles or Hollywood), which may indicate her locations. Here, we refer a *venue name* as the name for a geo signal, which could be a city (*e.g.*, Los Angeles), a place (*e.g.*, Time Square), or a local entity (*e.g.*, Stanford University). In the rest of the paper, we use "venue" for short. We note that a venue may refer different locations. *E.g.*, there are 19 towns named as "Princeton" in the States.

We formally abstract the two types of resources as "following" and "tweeting" relationships. A *following relationship*, denoted as $f\langle i, j\rangle$, is formed from a user $u_i$ to another user $u_j$ when $u_i$ follows $u_j$. $u_i$ is named as a *follower* of $u_j$, and $u_j$ is named as a *friend* of $u_i$. We use $f_{1:S}$ to represent all the following relationships, where $S$ is the total number of the relationships. A *tweeting relationship* $t\langle i,j\rangle$ is formed from a user $u_i$ to a venue $v_j$, if $u_i$ tweets $v_j$. As $u_i$ can tweet $v_j$ many times, there could be many tweeting relationships between $u_i$ and $v_j$. We use $t_{1:K}$ to represent all the tweeting relationships, where $K$ is the total number of the relationships.

Further, we assume a relationship is associated with the *location assignments* that the relationship is based on. Specifically, for $f\langle i,j\rangle$, the location assignments $x_i$ and $y_j$ indicate that $u_i$ follows $u_j$ as $u_i$ and $u_j$ are in $x_i$ and $y_j$, respectively. *E.g.*, Austin is the location assignments for both Carol and Lucy for their following relationship, which indicates that Carol follows Lucy as they were classmates in Austin. Similarly, for $t\langle i,j\rangle$ (*e.g.*, Carol tweets about "Hollywood"), the location assignment $z_i$ (Los Angeles) indicates $u_i$ (*e.g.*, Carol) tweets $v_j$ (*e.g.*, "Hollywood") because $u_i$ is interested in $z_i$. However, as a user's relationship could be related to any of her locations and its assignments are hidden to us, we need to profile its assignments.

Based on the above definitions, we formally abstract our problem as follows:

***User and Relationship Location Profiling*** Given a set of users $U$, which contains both labeled users $U^*$ and unlabeled users $U^N$, the home location $l_{u_i}$ for $u_i \in U^*$, their following and tweeting relationships $f_{1:S}$ and $t_{1:K}$, and candidate locations $L$, estimate a set of locations $\hat{L}_{u_i} \subset L$ for $u_i \in U$, location assignments $\hat{x}_i \in \hat{L}_{u_i}$ and $\hat{y}_j \in \hat{L}_{u_j}$ for $f\langle i,j\rangle \in f_{1:R}$, and a location assignment $\hat{z}_i \in \hat{L}_{u_i}$ for $t\langle i,j\rangle \in t_{1:K}$, so as to make $\hat{L}_{u_i}$, $\hat{x}_i$, $\hat{y}_j$ and $\hat{z}_i$ close to $u_i$'s location profile $L_{u_i}$ and the true assignments $x_i$, $y_j$ and $z_i$ respectively.

We note that the above problem estimates a set of locations for each user as well as location assignments for each relationships. The home location prediction problem studied by earlier work [8, 5, 11] can be viewed as its sub-problem, as we can estimate a user's home location as the most important location in the set. As discussed in Sec. 1, solving the problem is not easy and calls for a novel solution.

## 4. MULTIPLE LOCATION PROFILING

In this section, we develop *MLP* to profile locations for both users and their relationships with the following network and the tweeting content.

Our first goal is to connect the two types of relationships with users' locations. Intuitively, we can assume that both of them are "generated" based on a same set of latent variables — users' locations. Then, it naturally leads us to a *probabilistic generative approach*, which models the *joint probability* of generating the two types of relationships based on users' locations. We can estimate users' locations and location assignments for relationships as the latent variables in the probability.

However, as we have motivated in Sec. 1 and 2, to model the joint probability, we need to address the challenges of *location-based generation*, *mixture of observations* and *partially available supervision*, which have not been studied by the existing generative models like LDA and MMSB.

We propose *MLP* to model the joint probability and deal with those challenges. Fig. 2 shows its plate diagram and Tab. 1 gives notations. Generally, it illustrates how *MLP* models the joint probability that 1) generates each user $u_i$'s location distribution $\theta_i$ based on a hyper distribution with a parameter $\gamma_i$, which is determined by the observed locations from the labeled users, 2) generates location assignments (*e.g.*, $x_{s,i}$ and $z_{k,i}$) based on $\theta_i$, and 3) generates the associated following and tweeting relationships (*e.g.*, $f_s\langle i,j\rangle$ and $t_k\langle i,j\rangle$) based on the location assignments. Thus, we can estimate $\theta_i$, $x_{s,i}$, $y_{s,j}$ and $z_{k,i}$ with the observed relationships and locations, and use $\theta_i$ as $u_i$'s location profile.

Table 1: Notations

| | |
|---|---|
| $N$ | Total number of users |
| $L$ | All the candidate locations |
| $V$ | All the venue names |
| $\vec{\eta}_i$ | Observation vector for $u_i$ |
| $\vec{\lambda}_i$ | Candidacy vector for $u_i$ |
| $b_o, b_c$ | Bernoulli distributions that generate $\vec{\eta}_i$ and $\vec{\lambda}_i$ |
| $\Lambda$ | Boosting matrix |
| $\tau$ | Prior for candidate locations |
| $\theta_i$ | Location profile of $u_i$ |
| $\theta_{1:N}$ | Location profiles for $N$ users |
| $\gamma$ | General prior distribution parameter for $\theta_i$ |
| $\gamma_i$ | Prior distribution parameter for $\theta_i$ |
| $F_L, T_L$ | Location-based following and tweeting models |
| $\alpha, \beta$ | Parameters of $F_L$ |
| $\psi_l$ | Location-based tweeting model of $l$ |
| $\psi_{1:L}$ | Location-based tweeting models for $L$ |
| $T_R, F_R$ | Random tweeting and following models |
| $S$ | Total number of following relationships |
| $f_{1:S}$ | All the following relationships |
| $f_s\langle i,j\rangle$ | $s^{th}$ following relationship from $u_i$ to $u_j$ |
| $\mu_s$ | Model selector for $f_s\langle i,j\rangle$ |
| $\mu_{1:S}$ | Model selectors for $f_{1:S}$ |
| $x_{s,i}$ | Location assignment for $u_i$ in $f_s\langle i,j\rangle$ |
| $y_{s,j}$ | Location assignment for $u_j$ in $f_s\langle i,j\rangle$ |
| $x_{1:S}$ | Location assignments for followers in $f_{1:S}$ |
| $y_{1:S}$ | Location assignments for friends in $f_{1:S}$ |
| $K$ | Total number of tweeting relationships |
| $t_{1:K}$ | All the tweeting relationships |
| $t_k\langle i,j\rangle$ | $k^{th}$ tweeting relationship from $u_i$ to $v_j$ |
| $\nu_k$ | Model selector for $t_k\langle i,j\rangle$ |
| $\nu_{1:K}$ | Model selectors for $t_{1:K}$ |
| $z_{k,i}$ | Location assignment for $u_i$ in $t_k\langle i,j\rangle$. |
| $z_{1:K}$ | Location assignments for users in $t_{1:R}$. |

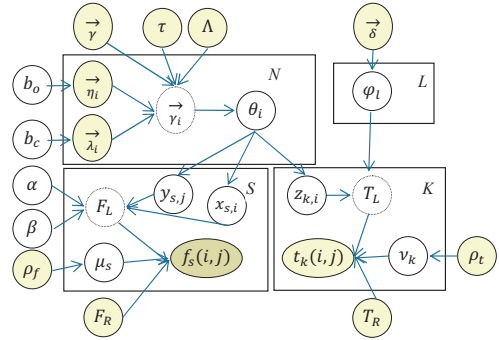

Figure 2: Plate Diagram for *MLP*

In the following parts, we first explain three key components of *MLP*, which deals with the above challenges, and then present *MLP* and its inference algorithm in detail.

### 4.1 Location-based Generation

We first present our location-based generative models, which formally measure the probability that a following or tweeting relationship (*e.g.*, $f\langle i,j\rangle$ or $t\langle i,j\rangle$) is generated given users' location assignments (*e.g.*, $x_i$, $y_j$ or $z_i$). In Fig. 2, they are represented by $F_L$ and $T_L$ respectively.

1606

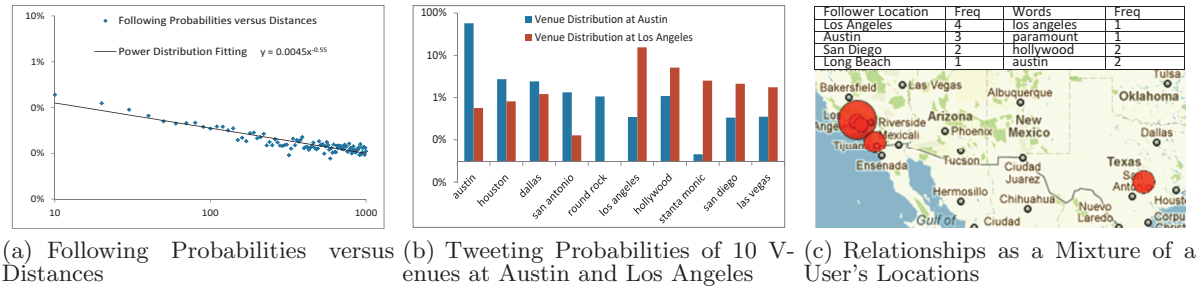

(a) Following Probabilities versus Distances  (b) Tweeting Probabilities of 10 Venues at Austin and Los Angeles  (c) Relationships as a Mixture of a User's Locations

Figure 3: Observations

The models should be carefully designed, as a user follows friends from different locations and tweets different venues. Fortunately, locations are predefined semantic attributes, and we observe locations and relationships of some users. Thus, we investigate a large-scale Twitter data (Sec. 5 gives the statistics of the data), and learn the models from there.

**Location-based Following Model** We begin with investigating the following probability of observing a following relationship $f\langle i,j\rangle$ from a user $u_i$ to a user $u_j$ given their locations $x_i$ and $y_j$. It involves two locations. If we view any pair of locations as simply two distinct categorical labels, we overlook the inherent relation between them. Thus, we explore the probability as a function of *distance*, since the distance is a natural and fine-grained measure for the relation between two locations.

Fig. 3(a) illustrates following probabilities over different distances. We first compute the distance between any pair of labeled users, resulting about $2.5 * 10^{10}$ pairs. Then we bucket them by intervals of 1 mile and measure the probability of generating a following relationship at $d$ miles as the ratio of the number of pairs that have following relationships to the total number of pairs in the $d^{th}$ bucket. We plot the probabilities versus distances in the log-log scale.

The figure shows that 1) the following probability decreases as the distance increases, and 2) at the distances in a long range, the probabilities do not decay as sharply as those at the distances in a short range. Such probabilities successfully capture our intuition that a user is likely to follow friends, who live close to him, but also may follow some users, who live far away. When he follows the users living far away, the following probabilities are less sensitive to his distances to them.

We can fit the probabilities in Fig. 3(a) with a power law distribution, as power laws are straight lines when they are plotted in the log-log scale. Mathematically, a power law distribution has two parameters, $\alpha$ and $\beta$, and the probability at a point $x$ is expressed as $P(x|\alpha,\beta) = \beta x^\alpha$. Given a set of observations, i.e., $x$ and $P(x|\alpha,\beta)$, we can learn $\alpha$ and $\beta$. In our case, $\alpha = -0.55$ and $\beta = 0.0045$.

Now, we formally describe our *location-based following model*. We model the following probabilities of whether there is $f\langle i,j\rangle$ from $u_i$ to $u_j$ given $x_i$ and $y_j$ as a Bernoulli distribution with a parameter $p$, and model $p$ at different distances $d(x_i,y_j)$ as a power law distribution with parameters $\alpha$ and $\beta$. Mathematically, we measure it as follows.

$$P(f\langle i,j\rangle|\alpha,\beta,x_i,y_j) = \beta d(x_i,y_j)^\alpha \quad (1)$$

We note that similar power law distributions have been observed in Facebook data [5] and other social networks [12], but this paper is the first study on Twitter and gives new observations. Specifically, the exponent is -0.55, which is different from -1 observed in the Facebook data [5]. It suggests that the following relationships on Twitter are less sensitive to users' distances than the friendships in Facebook. Therefore, profiling locations for Twitter users is more difficult than for Facebook users studied in [5]. It requires us to utilize additional resources and build an advanced model.

**Location-based Tweeting Model** Next, we explore the *tweeting probability* that a tweeting relationship $t\langle i,j\rangle$ is generated from a user $u_i$ to a venue $v_j$ given $u_i$'s location $z_i$. As a venue name (e.g., Princeton) may refer different locations (e.g., Princeton, NJ or Princeton, WV), we can not view it as a single location. Thus, we view venues as categorical labels and explore tweeting probabilities at a specific location as a discrete distribution over venues $V$.

Fig. 3(b) shows the tweeting probability of 10 venues by the users at Austin and Los Angeles. To generate Fig. 3(b), we first extract venues (city names) from users' tweets. Then, for each location, say Austin, we count the relative frequencies of the venues, and thus the probabilities, that the venues are tweeted by those users at the location. Due to the space limit, we only select the top five venues with the largest probabilities from each location, and plot their probabilities in the log scale.

We obtain the following observations. The tweeting probabilities of different locations are different over the same venues. E.g., users in Los Angeles are more likely to tweet "los angeles" than those in Austin. For tweeting probabilities at a location (e.g., Austin), we see that 1) nearby venues (e.g., "austin") have high probabilities to be tweeted, 2) far-away venues (e.g., "hollywood") have small probabilities to be tweeted, and 3) the probability to tweet a venue is not a monotonic function of its distance to the location. E.g., "hollywood" and "round rock" have similar probabilities to be tweeted by users in Austin, but Round Rock city is much closer than Hollywood. The tweeting probabilities so observed do reflect that users are likely to tweet their local venues as well as far but popular venues.

We develop our *location-based tweeting model* to capture the above observations. Specifically, for a location $l$, we use a multinomial distribution $\psi_l$ over venues $V$ to model the tweeting probabilities of $l$. $V$ can be defined based on a gazetteer. Each $l$ is associated with its own $\psi_l$, and there are totally $|L|$ multinomial distributions, denoted as $\psi_{1:L}$. We measure the tweeting probability that $u_i$ builds $t\langle i,j\rangle$ to $v_j$ given $z_i$ as the probability of picking $v_j$ from $\psi_{z_i}$. Mathematically, it is measured as follows.

$$P(t\langle i,j\rangle|\psi_{1:L},z_i) = P(v_j|\psi_{z_i}). \quad (2)$$

We note that the above distributions are obtained based on the locations provided by labeled users. The parameters (*e.g.*, $\alpha$ and $\beta$) in those distributions may not be precisely learned due to the noisy-signal and mixed-signal challenges, which will be discussed next. However, we believe the observations are reliable for choosing proper distributions to model the two probabilities. We can further precisely estimate those parameters as we will show in Sec. 4.5.

## 4.2 Mixture of Observations

To fundamentally deal with the noisy-signal and mixed-signal challenges motivated in Sec. 1, we introduce two level mixture components in *MLP*. The first level aims to capture that there are both "noisy" and "location-based" relationships, and the second level aims to address that the "location-based" relationships are related to users' multiple locations.

**The Noisy-signal Challenge** First, we argue that some relationships are not generated based on locations, and therefore are noises for profiling users' locations. *E.g.*, Carol in Austin follows Gaga in New York. We call those relationships as *noisy* relationships, and the remaining ones as *location-based* relationships. The previous methods [8, 5] do not model noisy relationships explicitly, and can not profile users' locations accurately.

We propose a mixture component to capture noisy and location-based relationships. Conceptually, we assume a relationship is generated based on either a *location-based generative model*, which is introduced above, or a *random generative model*, which we will introduce below. Technically, for each following relationship, we introduce a binary *model selector* $\mu$, where $\mu = 1$ means the random generative model is selected to generate the relationship, and 0 otherwise. We further assume that $\mu$ is generated based on a Bernoulli distribution with a parameter $\rho_f$, which models how likely a following relationship is generated based on the random generative model. Similarly, for a tweeting relationship, we introduce a model selector $\nu$ and a Bernoulli distribution with a parameter $\rho_t$ to generate $\nu$.

We now design the *random generative models*. Intuitively, we model the *random following model*, denoted as $F_R$, as a Bernoulli distribution to represent the probabilities of whether a following relationship is randomly built between two users. We model the *random tweeting model*, denoted as $T_R$, as a multinomial distribution over venues $V$ to represent the probabilities that a tweeting relationship is randomly built to venues from a user.

Similar to existing work [14], we learn $F_R$ and $T_R$ empirically. Specifically, we model $F_R$, which measures the probability that $u_i$ randomly builds $f\langle i,j\rangle$ to $u_j$, as $p(f\langle i,j\rangle = 1|F_R) = \frac{S}{N^2}$, where $S$ is the number of following relationships and $N^2$ is the total number of user pairs. We model $T_R$, which measures the probability that $u_i$ randomly builds $t\langle i,j\rangle$ to $v_j$, as $p(t\langle i,j\rangle|T_R) = \frac{\sum_{u_x \in U} t\langle x,j\rangle}{K}$, where $\sum_{u_x \in U} t\langle x,j\rangle$ is the number of tweeting relationships to $v_j$, and $K$ the total number of tweeting relationships.

**The Mixed-signal Challenge** Next, we argue that the location-based relationships are generated based on users' multiple locations. To illustrate, we give an example of the user with id 13069282. From the user's home page in her Twitter profile, we know that she used to study in Austin and now works in Los Angeles. Fig. 3(c) shows her friends' locations, tweeted venues, as well as a map with her friends' locations plotted. The figure clearly shows that her friends are in and her tweets are about the two regions, and suggests that a user follows friends from or tweet venues related to his multiple locations.

The previous methods [8, 5] haven't addressed this issue. They not only profile a user's locations incompletely, but also predict the home location incorrectly, because locations of the friends related to her other locations (*e.g.*, Austin) are noisy information to profile her home location (*e.g.*, Los Angeles). Although the our model can handle noises somehow, a lot of friends at great distances are "noisy" enough to make our model fail.

To fundamentally deal with the mixed-signal challenge, we first model a user $u_i$'s location profile as a multinomial distribution over candidate locations $L$, denoted as $\theta_i$. The probability of a location $l$ in $\theta_i$ represents how likely $u_i$ is at $l$. Our goal is to estimate $\theta_i$ for each $u_i$. We then assume that a location-based relationship is generated based on a specific location assignment picked from each related user's profile, rather than their home locations only.

Thus, we extend our location-based models into two stage *generative processes*. Specifically, the *location-based following process* models that a location-based following relationship $f\langle i,j\rangle$ from $u_i$ to $u_j$ is generated via the following two steps: 1) randomly select two location assignments $x_i$ and $y_j$ from $\theta_i$ and $\theta_j$, and 2) randomly generate $f\langle i,j\rangle$ based on the location-based following model $F_L$, specifically, $P(f\langle i,j\rangle|x_i,y_j,\alpha,\beta)$. Similarly, the *location-based tweeting process* models that a tweeting relationship $t\langle i,j\rangle$ from $u_i$ to $v_j$ is generated via the following two steps: 1) random select a location $z_i$ from $\theta_i$, and 2) randomly generate $t\langle i,j\rangle$ based on the location-based tweeting model $T_L$, specifically, $P(t\langle i,j\rangle|z_i,\phi_{z_i})$.

We note that the location assignments for a relationship explain the true geo connection in the relationship in terms of users' hidden locations rather than users' home locations only, and thus help us to fundamentally capture that a user's relationships are generated based on her multiple locations.

## 4.3 Partially Available Supervision

To incorporate home locations from labeled users as supervision, we further model how a user's location profile $\theta_i$ is generated by a prior distribution with a particularly derived parameter, denoted as $\gamma_i$ in the plate diagram.

First, we motivate the need for supervision. By far, our model runs in an "unsupervised" way as LDA and MMSB. It assumes that relationships are generated based on users' location profiles, and can estimate them with the relationships. It neither models nor requires that locations of some users are observed. However, without an "anchoring" point, which is known somehow, the hidden clusters of "near locations" would be floating. For example, given a set of densely connected users, our model can tell that they are likely in a location, but can not identify which location (*e.g.*, Los Angeles or Austin) they are in. In reality, 16% Twitter users provide their home locations. If our model captures some of the users in the example are in Los Angeles, it can accurately learn location profiles for all of them.

However, there is no obvious way of incorporating observed locations as supervision. First, we can not set a user's $\theta_i$ as observed, because we observe only his home location instead of his location profile. Second, we can not

1608

set the location assignments for his relationships as the observed location, as it does not allow the relationships to be generated based on other locations and fails to address the mixed-singal challenge. The existing modifications of LDA incorporate supervision in different settings. For example, the supervised LDA model [6] assumes a document has a label and each label corresponds to a mixture of topics. Our setting is different. First, we view each hidden dimension (a topic in LDA) as a sematic label (location). Second, a user has multiple labels, but only one label is observed.

We choose to use the home locations of labeled users as prior knowledge to generate their location profiles. As LDA, we assume that a user's location profile $\theta_i$ is generated from a Dirichlet distribution $\text{DIR}(\vec{\gamma})$ with a hyper parameter $\vec{\gamma}$. In $\text{DIR}(\vec{\gamma})$, the larger $\vec{\gamma}$'s $l^{th}$ dimension $\gamma_l$ is, the more likely $\theta_i$ with a large probability in the $l^{th}$ dimension is to be generated. However, in LDA, $\vec{\gamma}$ is set uniformly, as it does not have any preference on any topic, while we can set them differently to encode our prior knowledge for labeled users, as we observe their home locations.

Technically, we introduce an "observation vector" and a "boosting matrix" to set the prior for each user. For a user $u_i$, an *observation vector* is an L-length binary vector, denoted as $\vec{\eta}_i$, and its $j^{th}$ dimension $\eta_{i,j}$ represents whether the $j^{th}$ location is observed. We assume $\eta_{i,j}$ is generated via a Bernoulli process with a parameter $b_o$, but is observed. A *boosting matrix* is an $L \times L$ matrix, denoted $\Lambda$, and a cell $\Lambda_{ij}$ represents how much the prior of the $j^{th}$ location should be boosted when the $i^{th}$ location is observed. In our implementation, we assume $\Lambda$ is a diagonal matrix for simplicity, which means observing the $i^{th}$ location only boots its prior. Thus, the hyper parameter $\vec{\gamma}_i$ for $u_i$ is set by $\vec{\gamma}_i = \vec{\eta}_i \times \Lambda \times \vec{\gamma} + \vec{\gamma}$, where the first term encodes how much we boost the prior for an observed location, and the second term encodes our priors for candidate locations. With $\vec{\gamma}_i$, we will have a high probability to obtain $\theta_i$ that has a high probability to generate the observed location. We will see this clearly in Sec. 4.4.

Then, we motivate the need for limiting the number of candidate locations in a user's location profile. There are three reasons. First, it is useless to consider every location for a user, as some are definitely not related to him. *E.g.*, if a user only follows users in and tweets about California, any location from the east coast is not related to him. Second, a user usually has a small number of locations due to relocation costs. Third, it is inefficient to consider every location for every user. We will show this clearly in Sec. 4.5.

This is a unique challenge in our setting and has not been addressed by LDA, because in LDA the number of topics can be adjusted (usually from 20 to 200) during the estimation, while in *MLP*, a set of candidate locations $L$ is given, which could be a very large number (5000 in our experiment).

To solve the challenge, we introduce a "candidacy vector" to represent the candidacy of locations for a user $u_i$. For $u_i$, his *candidacy vector* is an L-length binary vector, denoted as $\vec{\lambda}_i$. $\lambda_{i,j}$ is 1 if and only if the $j^{th}$ location is a candidate location for $u_i$. We can assume $\lambda_{i,j}$ is generated via a Bernoulli process with a parameter $b_c$, but is observed.

We utilize location observed from a user's neighbors to set his candidacy vector. Specifically, we assume that $\vec{\lambda}_{i,j}$ is 1, if and only if the $j^{th}$ candidate location is observed from $u_i$'s following and tweeting relationships. The statistics from our data generally validate this assumption. In our incomplete crawl of Twitter, there are about 92% users whose locations appear in their relationships. We use $\tau$ to represent the prior value for each candidate location. $\tau$ is set to a small number (0.1 in our experiments), as previous studies show [7] that the values of hyper parameter below 1 prefer sparse distributions. Thus, we can use $\tau \cdot \vec{\lambda}_i$ to represent priors of candidates locations for $u_i$.

Thus, the prior $\gamma_i$ for a user $u_i$ can be set as follows,

$$\vec{\gamma}_i = \vec{\eta}_i \times \Lambda \times \vec{\gamma} + \tau \cdot \vec{\lambda}_i. \quad (3)$$

### 4.4 Generative Model

We now present *MLP* completely. As a generative model, it can be explained by an imaginary process that describes how following and tweeting relationships are generated.

**Generative Process** First, for each user $u_i$, we generate his prior distribution parameter $\gamma_i$ and location profile $\theta_i$.

- Generate $u_i$'s observation vector $\vec{\eta}_i$ via a Bernoulli distribution with a parameter $b_o$.
- Generate $u_i$'s candidacy vector $\vec{\lambda}_i$ via a Bernoulli distribution with a parameter $b_c$.
- Calculate $\vec{\gamma}_i$ based on Eq. 3.
- Generate $\theta_i$ from a Dirichlet distribution with $\vec{\gamma}_i$.

We note that since $\vec{\eta}_i$ and $\vec{\lambda}_i$ are observed, they block the influence of $b_o$ and $b_c$. We can ignore $b_o$ and $b_c$ in the joint probability. As $\vec{\gamma}_i$ can be computed from $\vec{\eta}_i$ and $\vec{\lambda}_i$, we will use the computed $\gamma_i$ in the joint probability directly.

Second, for each location $l$, its tweeting model $\psi_l$ is generated from a Dirichlet distribution $\text{DIR}(\vec{\delta})$.

Third, for each pair of users $u_i$ and $u_j$, whether $u_i$ builds a following relationship $f\langle i, j\rangle$ to $u_j$ is determined as follows.

- Generate a model selector $\mu$ according to a Bernoulli distribution with a parameter $\rho_f$.
- If $\mu = 1$, we choose the random following model $F_R$ to decide whether there is $f\langle i, j\rangle$.
- if $\mu = 0$, we choose the location-based following process, which contains the following steps.
- Choose a location assignment $x_i$ from $\theta_i$.
- Choose a location assignment $y_j$ from $\theta_j$.
- Decide whether there is $f\langle i, j\rangle$ based on the location-based following model as shown in Eq. 1.

We note that the above process models any pair of users including pairs with or without a following relationship. However, we choose to use only the pairs with following relationships as our observations because of two reasons. First, it is more faithful to the underlying semantics of the data in our setting, as the absence of a following relationship from $u_i$ to $u_j$ does not necessarily mean that $u_i$ will not follow $u_j$. *E.g.*, they may be real friends who are unaware of each other's existence in the network. Second, it significantly decreases the computational cost of inference, as the complexity of computation scales with the number of observed relationships rather than the number of user pairs.

Fourth, for each tweeting relationship $t_k \langle i, j\rangle$ from a user $u_i$ to a venue $v_j$, it is generated by the following steps.

- Generate a model selector $\nu_k$ according to a Bernoulli distribution with a parameter $\rho_t$.

- If $\nu_k = 1$, we choose the random tweeting model $T_R$ to generate $t_k\langle i,j\rangle$.
- If $\nu_k = 0$, we choose the location-based generation process, which contains the following steps.
- Choose a location assignment $z_{k,i}$ from $\theta_i$.
- Generate $t_k\langle i,j\rangle$ based on the location-based tweeting model as shown in Eq. 2.

**Joint Probability** Based on the generative process, *MLP* defines the *join probability* of generating both the observed and hidden random variables given model parameters. Specifically, we assume the parameters, $\rho_f$, $\rho_t$, $\alpha$, $\beta$, $F_R$, $T_R$, $\vec{\gamma}_i$ and $\vec{\delta}$ are given. To simplify our notations, we use $\Omega$ to represent them. The joint distribution can be represented as follows.

$$P(\theta_{1:N}, \psi_{1:L}, \mu_{1,S}, x_{1:S}, y_{1:S}, f_{1:S}, \nu_{1:K}, z_{1:K}, t_{1:K}|\Omega)$$
$$= \prod_{i=1}^{N} P(\theta_i|\vec{\gamma}) \prod_{l=1}^{L} P(\psi_l|\vec{\delta}) \prod_{k=1}^{K} P(\nu_k|\rho_t) \prod_{s=1}^{S} P(\mu_s|\rho_s)$$
$$\prod_{s=1}^{S} (P(x_{s,i}|\theta_i)P(y_{s,j}|\theta_j)P(f_s\langle i,j\rangle|\alpha,\beta,x_{s,i},y_{s,j}))^{1-\mu_s}$$
$$\prod_{k=1}^{K} (P(z_{k,i}|\theta_i)P(t_k\langle i,j\rangle|z_{k,i}=l,\psi_l))^{1-\nu_k}$$
$$\prod_{s=1}^{S} P(f_s\langle i,j\rangle|F_R)^{\mu_s} \prod_{k=1}^{K} P(t_k\langle i,j\rangle|T_R)^{\nu_k} \quad (4)$$

In the above equation, the following and tweeting relationships, i.e., $f_{1:S}$ and $t_{1:K}$, are observed, while users' location profiles $\theta_{1:N}$, the locations' tweeting models $\psi_{1:L}$, the model selectors (*e.g.*, $\mu_s$, $\nu_k$) and the location assignments (*e.g.*, $x_{s,i}$ $y_{s,j}$ and $z_{k,i}$) are hidden. The central computational problem for *MLP* is to use the observed relationships and the given parameters to infer the hidden unknown variables.

**Discussions** Based on Fig. 2, we can clearly explain the difference between *MLP* and MMSB mentioned in Sec. 2 in terms of generating "relationships" between nodes based on pairwise variable interactions. MMSB associates every pair of communities with an interaction parameter and uses $K^2$ parameters for $K$ communities, while *MLP* uses a power law distribution with $\alpha$ and $\beta$ to parameterize pairwise location interactions based on the real-world observations in Sec. 4.1. *MLP* has two advantages. First, it greatly reduces the number of parameters from $K^2$ to 2, and thus parameters can be estimated accurately with limited observations. Second, it explicitly constrains "interaction probabilities" and makes location profiling accurate. The interaction probabilities in MMSB could be any distribution, while the power law distribution explicitly constraints that the two users in a relationship are likely to be close.

### 4.5 Inference with Gibbs Sampling

*MLP* models various aspects that haven't been addressed by existing generative models, and combines discrete and continuous distributions in a non-trivial manner. It is complex and does not allow for exact inference. We derive our own approximate inference algorithm.

Specifically, we derive our inference algorithm via the following steps: 1) we integrate $\theta_{1:N}$ and $\psi_{1:L}$ in the joint probability, so we do not need to estimate $\theta_{1:N}$ and $\psi_{1:L}$ at the beginning, 2) we use the Gibbs sampling method, which is one of classical sampling methods, to sample from the posterior distribution of the model selectors and the location assignments given the relationships and the model parameters, $P(\mu_{1,S}, x_{1:S}, y_{1:S}, \nu_{1:K}, z_{1:S}|f_{1:S}, t_{1:K}, \Omega)$, and 3) we estimate the location profile $\theta_i$ for each user $u_i$ based on sampled $\mu_{1,S}$, $\nu_{1,K}$, $x_{1:S}, y_{1:S}$ and $z_{1:K}$.

To sample from $P(\mu_{1,S}, x_{1:S}, y_{1:S}, \nu_{1:K}, z_{1:S}|f_{1:S}, t_{1:K}, \Omega)$, a standard Gibbs sampling procedure requires to compute the following conditional posterior distributions.

- $P(\mu_s|\mu_{-s}, \nu_{1:S}, x_{1:S}, y_{1:S}, f_{1:S}, z_{1:K}, t_{1:K}, \Omega)$,
- $P(\nu_k|\nu_{-k}, \mu_{1:S}, x_{1:S}, y_{1:S}, f_{1:S}, z_{1:K}, t_{1:K}, \Omega)$,
- $P(x_{s,i}|\mu_{1:S}, \nu_{1:S}, x_{-s:i}, y_{1:S}, f_{1:S}, z_{1:K}, t_{1:K}, \Omega)$,
- $P(y_{s,j}|\mu_{1:S}, \nu_{1:S}, x_{1:S}, y_{-s:j}, f_{1:S}, z_{1:K}, t_{1:K}, \Omega)$,
- $P(z_{k,i}|\mu_{1:S}, \nu_{1:S}, x_{1:S}, y_{1:S}, f_{1:S}, z_{-k:i}, t_{1:K}, \Omega)$,

In the above probabilities, $\mu_{-s}$, $\nu_{-k}$, $x_{-s,i}$, $y_{-s,j}$, or $z_{-k,i}$ denote all the assignments except the $s^{th}$ or $k^{th}$ assignment. We derive those equations as below. The detailed derivation is omitted due to the space limitation.

$$P(\mu_s|\mu_{-s}, \nu_{1:S}, x_{1:S}, y_{1:S}, f_{1:S}, z_{1:K}, t_{1:K}, \Omega)$$
$$\sim P(\mu_s|\rho_f)(P(f_s\langle i,j\rangle|F_R))^{\mu_s} \times$$
$$(\frac{\varphi_{i,l} + \gamma_{i,l} - 1}{\varphi_i + \sum_{l=1}^{L}\gamma_{i,l} - 1}\beta \times d(x_{s,i}, y_{s,j})^{\alpha})^{1-\mu_s} \quad (5)$$

$\varphi_{i,l}$ denotes the frequency that the $l^{th}$ location has been observed from $u_i$'s location assignments. $\varphi_i$ denotes the total number of $u_i$'s location assignments. $\gamma_{i,l}$ is the $l^{th}$ dimension of the prior $\vec{\gamma}_i$.

$$P(\nu_k|\nu_{-k}, \mu_{1:S}, x_{1:S}, y_{1:S}, f_{1:S}, z_{1:K}, t_{1:K}, \Omega)$$
$$\sim P(\nu_k|\rho_t)(P(t_s\langle i,j\rangle|T_R))^{\nu_k} \times$$
$$(\frac{\varphi_{i,l} + \gamma_{i,l} - 1}{\varphi_i + \sum_{l=1}^{L}\gamma_{i,l} - 1}\frac{\phi_{l,v} + \delta_v - 1}{\sum_{v=1}^{V}(\phi_{l,v} + \delta_v) - 1})^{1-\nu_k} \quad (6)$$

$\phi_{l,v}$ is the frequency that $v$ is tweeted by users at $l$. $\delta_v$ is the vth dimension of the prior $\vec{\delta}$.

The above two equations sample model selectors of relationships, which help us to identify noisy relationships. They can be interpreted intuitively. For example, in Eq. 5, the probability of $\mu_s = 1$ is proportional to two factors: 1) the probability of $\mu_s = 1$ encoded in $\rho_f$, and 2) the probability of observing $t_s\langle i,j\rangle$ in the random model $F_R$.

$$P(x_{s,i}|\mu_{1:S}, \nu_{1:S}, x_{-s:i}, y_{1:S}, f_{1:S}, z_{1:K}, t_{1:K}, \Omega)$$
$$\sim \frac{\varphi_{i,l} + \gamma_{i,l} - 1}{\varphi_i + \sum_{l=1}^{L}\gamma_{i,l} - 1}(d(x_{s,i}, y_{s,j})^{\alpha})^{1-\mu_s} \quad (7)$$

$$P(y_{s,j}|\mu_{1:S}, \nu_{1:S}, x_{1:S}, y_{-s:j}, f_{1:S}, z_{1:K}, t_{1:K}, \Omega)$$
$$\sim \frac{\varphi_{j,l} + \gamma_{j,l} - 1}{\varphi_l + \sum_{l=1}^{L}\gamma_{j,l} - 1}(d(x_{s,i}, y_{s,j})^{\alpha})^{1-\mu_s} \quad (8)$$

$$P(z_{k,i}|\mu_{1:S}, \nu_{1:S}, x_{1:S}, y_{1:S}, f_{1:S}, z_{-k:i}, t_{1:K}, \Omega)$$
$$\sim \frac{\varphi_{i,l} + \gamma_{i,l} - 1}{\varphi_i + \sum_{l=1}^{L}\gamma_{i,l} - 1}(\frac{\phi_{l,v} + \delta_v - 1}{\sum_{v=1}^{V}(\phi_{l,v} + \delta_v) - 1})^{1-\nu_k} \quad (9)$$

The above three equations sample location assignments for relationships, which can be viewed the estimated location assignments that explain the true geo connections in the relationships. They can be interpreted intuitively. For example, Eq. 7 contains two parts. The first one suggests that the probability of $x_{s,i} = l$ should be proportional to the frequency of the $l^{th}$ location in the existing samples of $u_i$ plus our prior belief $\gamma_l$. The second one suggests that the probability should be negatively related to the distance from $x_{s,i}$ to $y_{s,j}$ (remind that $\alpha$ is learned as $-0.55$ initially), but this part is active when the location-based model is used ($\mu_s = 0$). When the random model is used ($\mu_s = 1$), the probability is only proportional to the first part.

Our algorithm performs the above update equations for every following and tweeting relationship in one iteration. The algorithm runs a number of iterations until convergence.

From the above equations, we can clearly see that the supervision is encoded in our model. $\gamma_{i,l}$ can be interpreted as pseudocounts for the $l^{th}$ location in $\theta_i$. Remind that we set $\gamma_{i,l}$ high when the $l^{th}$ location is observed from the $i^{th}$ user. Thus, we will have a high probability to generate the observed location for a labeled user.

From the above equations, we can also see that users' candidacy vectors greatly improve the efficiency our algorithm. As Eq. 7, 8 and 9 estimate a probability for each candidate location for each assignment, the candidacy vector helps us to prune a large set of unrelated locations, and we do not need to estimate their probabilities.

After obtaining the location assignments for relationships, we estimate the location distribution $\theta_i$ for user $u_i$ with the maximal likelihood estimation principle.

$$p(l|\theta_i) = \frac{\varphi_{i,l} + \gamma_{i,l}}{\varphi_i + \sum_{l=1}^{L} \gamma_{i,l}} \quad (10)$$

Given the estimated $\theta_i$, we can predict $u_i$'s the home location as the one with the largest probability in $\theta_i$, and $u_i$'s location profile as the top K locations in $\theta_i$ or the locations whose probabilities are larger than a threshold.

Furthermore, we can apply the Gibbs-EM principle [3] to refine $\alpha$ and $\beta$ in our model. Specifically, at the E-step, we use the same Gibbs sampling algorithm to estimate $x_{s,i}$ and $y_{s,i}$'s distribution and calculate the expected distance of each following relationship. At the M-step, we estimate $\alpha$ and $\beta$ based on the expected distance for each following relationship. Therefore, the new algorithm contains two iterations. In the inner iteration, it uses Eq. 7, 8 and 9 to estimate the location assignments iteratively. The outer iteration computes $\alpha$ and $\beta$ iteratively according to the results from the inner iteration.

## 5. EXPERIMENT

In this section, we conduct extensive experiments on a large-scale data set to demonstrate the effectiveness of our model. Specifically, we first evaluate our model on the home location prediction task, and demonstrate that our model predicts users' home locations accurately and improves two state-of-the-art methods significantly. We further evaluate our model on discovering users' multiple locations and explaining following relationships, and show our model discover users' multiple locations completely and makes an accurate explanation for each relationship.

**Data Collection** We constructed our data set by crawling Twitter. We randomly selected 100,000 users as seeds to crawl in May 2011. For each user, we crawled his profile, followers and friends. After crawling, we obtained 3,980,061 users' profiles and their social network. Then, we extracted their registered locations from their profiles based on the rules described in [8]. Specifically, we extracted locations with city-level labels in the form of "cityName, stateName" and "cityName, stateAbbreviation," where we considered all cities listed in the Census 2000 U.S. Gazetteer. We found 630,187 users with city level locations and treated them as labeled users. Among them, we found 158,220 users, who had at least one labeled friend or follower. We crawled their tweets and extracted venues from them based on the same gazetteer. We crawled at most 600 tweets for each user. As we could not get some users' tweets due to their privacy settings or lack of tweets, only 139,180 users' tweets were crawled. We used the 139,180 users as well as their relationships and tweets, as our data set. There are 14.8 friends, 14.9 followers, and 29.0 tweeted venues per user.

**Tasks** We evaluate our model's performance on three tasks. Specifically, we apply our model to profile users' locations, and evaluate it on two tasks: 1) *home location prediction* and 2) *multiple locations discovery*. Then, we evaluate our model for *explaining following relationships*.

**Methods** To demonstrate the effectiveness of our model, we not only compare our model with two state-of-art methods in [5] and [8], but also evaluate our model with different types of resources. Specifically, we evaluate the following methods.

- $Base_U$ is the method in [5], which predicts a user's location based on his social network.
- $Base_C$ is the method in [8], which classifies a user into locations based on local words identified from tweets.
- $MLP_U$ is our prediction method, but only uses users' following relationships as observations.
- $MLP_C$ is our prediction method, but only uses users' tweeting relationships as observations.
- $MLP$ is our method discussed in Sec. 4, which uses both following and tweeting relationships as observations.

### 5.1 Results for Home Location Prediction

We first present our experiment results for predicting users' home locations.

**Ground Truth** To get users' home locations, we took their registered locations as their home locations, and applied five fold validation, which means that we used 80% of users as labeled users and 20% of users as unlabeled users and reported our results based on the average of 5 runs. We note that we directly took users' registered locations as their home locations, because we wanted to set up our experiments in the same way as the existing methods [8, 5]. We are aware that some registered locations are incorrect, but we believe they are rare, as leaving profiles empty is always an easy option. Therefore, our results are reliable overall.

**Measures** To evaluate performance, we applied *Accuracy within m miles* ($ACC@m$) used in [8] and [5] as our measure. Particularly, for a user $u$, let $l_u$ be $u$'s home location, $\hat{l}_u$ be the predicted one, and $d(l_u, \hat{l}_u)$ be their distance. For a set of test users $U$, $ACC@m = \frac{|\{u_i | u_i \in U \wedge d(l_u, \hat{l}_u) \leq m\}|}{|U|}$. By default, we set $m$ to 100.

Table 2: Home Location Prediction Results

| Method | $Base_U$ | $Base_C$ | $MLP_U$ | $MLP_C$ | $MLP$ |
|---|---|---|---|---|---|
| ACC@100 | 52.44% | 49.67% | 58.8% | 55.3% | **62.3%** |

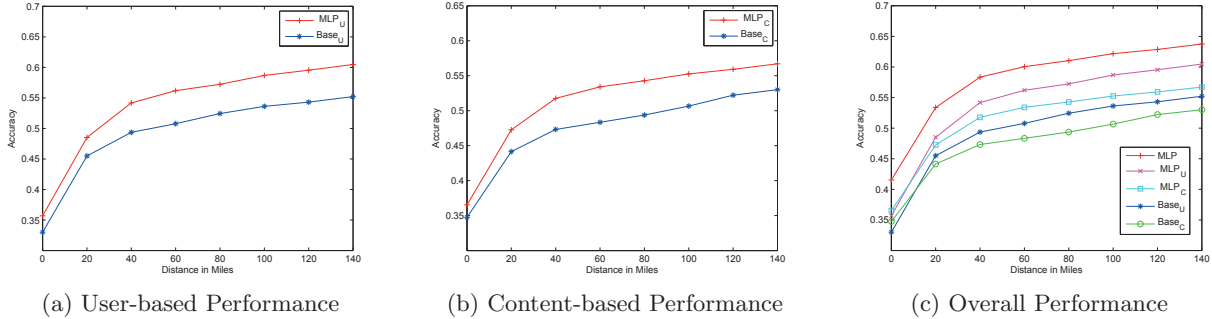

(a) User-based Performance  (b) Content-based Performance  (c) Overall Performance

Figure 4: Accumulative Accuracy at Various Distance

**User-based Performance** First, we compare $MLP_U$ with $Base_U$. Both of them profile a user's location based on his social network. Tab. 2 shows the results. $MLP_U$ improves $Base_U$ by 6% in terms of $ACC@100$. To illustrate the results in detail, we plot an *accumulative accuracy at distances* ($AAD$) curve for each method in Fig. 4(a). A point $(X,Y)$ in the curve means that Y percentages of users are accurately predicted within $X$ miles. From the figure, we can tell that $MLP_U$ has higher accuracy than $Base_U$ at different distances. E.g., $MLP_U$ places about 49% of users within 20 miles, while $Base_U$ only places 44% of users within that range. We believe the improvement results from explicitly dealing with the noisy-signal and mixed-signal challenges. Although we predict the home location of a user, modeling multiple locations helps us to rule out "noisy" following relationships and make accurate predictions.

**Content-based Performance** Next, we compare $MLP_C$ with $Base_C$. Both of them profile a user's location with his tweets only. From the results in Tab. 2 and the $AAD$ curves in Fig.4(b), we can clearly see that 1) $MLP_C$ significantly improves $Base_C$ by 5% in terms of $ACC@100$, and 2) the improvement is consistent at any distance level. Thus, we conclude that $MLP_C$ is better than $Base_C$, and we believe the improvement is due to explicitly modeling users' multiple locations and noisy venues.

We clarify that $Base_C$ requires human labeling to train a model to select local words, which are used as features for the classification model, and $Base_C$'s performance highly depends on the selected words. As the labeling is a subjective task, by no means could we get the same set of local words as in the original paper. We test performances of $Base_C$ with various local word sets, and we get $ACC@100$ ranging from 35.98% to 49.67%. We choose the highest one to report. Our method advances $Base_C$ in this aspect, as we do not require any labeling work, and only use venue names in an existing gazetteer.

**Overall Performance** Then, we compare $MLP$ with $Base_U$, $Base_C$, $MLP_U$, and $MLP_C$. Tab. 2 shows that $MLP$ improves the best baseline method $Base_U$ by 10%, and advances $MLP_C$ and $MLP_U$ by 7.0% and 3.6% respectively. Fig. 4(c) shows that those improvements are consistent at any distance level. We conclude that integrating different types of resources is useful, and our model can integrate them in a meaningful way. Meanwhile, we can say $MLP$ is very accurate. It correctly places 54% of users within 20 miles, and 62% users within 100 miles.

**Convergence** We also evaluate the convergence of our model. Fig. 5 shows the convergence rounds of $MLP$. It converges quickly after about 14 rounds of iterations. We note that the number of iterations is much less than other cases where the Gibbs sampling algorithm is applied (e.g., hundreds iterations in LDA [16]). We believe that our model converges quickly because we initialize each user's candidate locations based on our observations as discussed in Sec. 4.3.

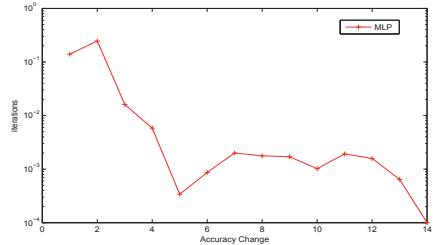

Figure 5: Accuracy Change in 14 Iterations

### 5.2 Results for Multiple Location Discovery

We continue our evaluations to see whether our model can capture and discover users' multiple locations.

**Ground Truth** To evaluate our model for discovering users' multiple locations, we first got the ground truth. As a user's profile does not contain multiple locations, we manually labeled locations for 1,000 users of the 139,180 users, and obtained 585 users, who clearly have multiple locations. We used those 585 users to evaluate our model and baseline methods. On average, a user has 2 locations.

To label users' related locations, we explored different sources. The first one is user profiles. Some profiles explicitly state multiple locations (e.g., Augusta, GA/New London, CT), or contain external links (e.g., linkedin accounts), which provide detailed information. The second one is tweets. Some tweets clearly express the user's related locations (e.g., "praying for my hometown. houston is wilding out."), and some contain GPS tags. Our labeling requirements are very strict. We do not consider a location as a related location for a user, if it just appears several times in his tweets but does not indicate that the user lives or lived there (e.g., "watching houston game").

**Measures** To evaluate the results, we introduce two new measures, *distance-based precision* ($DP$) and *distance-based recall* ($DR$). Specifically, we want to evaluate whether a set of discovered locations is close to a set of related locations of a user. In information retrieval, precision and recall evaluate whether retrieved results are relevant to a set of answers. However, they may underestimate performances in this task, because a predicted location (e.g., Santa Monica) may be different from but fairly close to a true location

1612

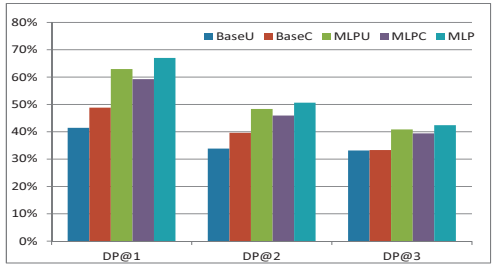

Figure 6: DP at Different Ranks

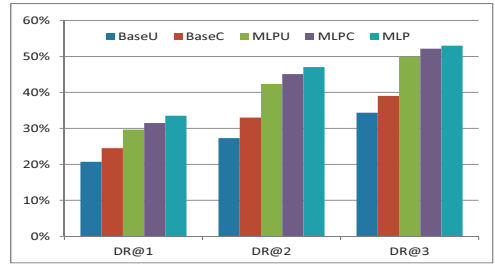

Figure 7: DR at Different Ranks

(*e.g.*, Beverly Hills). Therefore, we propose *DP* and *DR*. Intuitively, *DP* is the fraction of predicted locations that are close enough to true locations, while *DR* is the fraction of true locations that are close enough to predicted ones. Formally, we define that a location $l$ is close enough to a set of locations $L$, denoted as $c(l, L) = true$, if and only if $\exists l' \in L, s.t., D(l, l') < m$, where $m$ is a threshold and is set to 100 miles. For a user $u$, let $L'(u)$ and $L(u)$ be predicted and true locations for $u$. $DP(u) = \frac{|\{l | l \in L'(u) \wedge c(l, L(u))\}|}{|L'(u)|}$ and $DR(u) = \frac{|\{l | l \in L(u) \wedge c(l, L'(u))\}|}{|L(u)|}$. To measure *DP* and *DR* for a set $U$ of users, we average $DP(u)$ and $DR(u)$ for $u \in U$. We use *DP@K* or *DR@K* to denote *DP* or *DR* of the top K results. K is set to 2 by default, as users have 2 locations on average. As $Base_U$ and $Base_C$ find only one location, we use their top K predicted locations as the related locations.

Table 3: Multiple Location Discovery Results

| Method | $Base_U$ | $Base_C$ | $MLP_U$ | $MLP_C$ | $MLP$ |
|---|---|---|---|---|---|
| DP@2 | 33.8% | 39.3% | 45.1% | 48.3% | 50.6% |
| DR@2 | 27.2% | 33.1% | 42.3% | 45.3% | 47.0% |

**Overall Performance** Tab. 3 shows the performance of each method. Generally, our methods, $MLP_U$, $MLP_C$ and *MLP*, perform better than the baselines in both measures. In terms of *DP@2*, our methods predict more accurately than the baseline methods. In terms of *DR@2*, our methods discover users' locations more completely than the baseline methods. We believe that such advantages are achieved because our model fundamentally captures that a user has multiple locations. For example, when a user has multiple locations from different areas, our methods discovers them completely, while the baseline methods retrieve only one location and its nearby cities.

In addition, we plot *DP* and *DR* at different ranks in Fig. 6 and 7. From the figures, we obtain the following observations. First, our methods are better than the baseline methods at every $K$. Second, recalls (from *DR@1* to *DR@3*) of the baseline methods do not increase as much as those of our methods, when $K$ increases. It indicates that the baseline methods are not good at discovering multiple locations. Third, if we look at *DP@1*, baseline methods perform much worse than our methods. It is because when a user has multiple locations, his relationships generated based on other locations are noisy information for the baseline methods. It again validates that a user's multiple locations should be captured even for profiling his home location.

**Case Studies** To illustrate the correctness of our model in discovering multiple locations, we give some examples in Tab. 4. It clearly shows that our model finds multiple locations completely and accurately, while the baseline methods find only one of true locations and its nearby locations. For instance, for user 13069282 in the 2nd row, who studied in Austin and works in Los Angeles, *MLP* discovers both locations, while the top 2 results returned by $Base_U$ are all around Los Angeles area.

Table 4: Case Study on Multiple Location Discovery

| UID | True Locations | $MLP$ | $Base_U$ |
|---|---|---|---|
| 1178-4102 | St. Louis, MO<br>Anaheim, CA | St. Louis, MO<br>Los Angeles, CA | St. Louis, MO<br>Chicago, IL |
| 1306-9282 | Los Angeles, CA<br>Austin, TX | Los Angeles, CA<br>Austin, TX | Los Angeles, CA<br>San Diego, CA |
| 1501-3125 | Nashville, TN<br>Chicago, IL | Murfreesboro, TN<br>Chicago, IL | New York, NY<br>Franklin, TN |

### 5.3 Results for Relationship Explanation

We further evaluate our model to see whether relationships are correctly profiled.

**Ground Truth** To get the location assignments in following relationships, we manually labeled following relationships of the 585 users, whose multiple locations are known to us. In the labeling process, we only kept the following relationships in which users' location assignments could be clearly identified by their shared "regions" (*e.g.*, a user at Hollywood follows Los Angeles Weather Channel), and we obtained 4,426 relationships and the location assignments of them.

**Measure** We use *Accuracy within m miles* (*ACC@m*) as our measure. We define that a relationship is accurately explained if and only if both users' locations in the relationship are accurately assigned within $m$ miles.

As no previous work assigns locations for a relationship, we design a home location based explanation method to compare, denoted as *Base*. Specifically, for a following relationship, it directly assigns users' home locations as their location assignments in the relationship. It is a strong baseline, as users are likely to follow others based on their home locations, and in most cases we do not know users' home locations. However, this method will not work for the cases where users follow others based on their other locations.

**Overall Performance** Fig. 8 shows the *ACC@m* of each method with different $m$. Generally, we see *MLP* is significantly better than *Base*. Specifically, *Base* profiles only 40% relationships correctly. It again validates our assumption that a user's following relationships are not necessarily generated based on his home location. *MLP* significantly improves *Base* by 15%, which suggests that *MLP* correctly profiles each relationship and so as to profile users' locations accurately. The advantages are consistent with different distances. However, *ACC@50* of *MLP* is almost the same as *ACC@100*, which means most of the correctly profiled relationships are profiled within 50 miles.

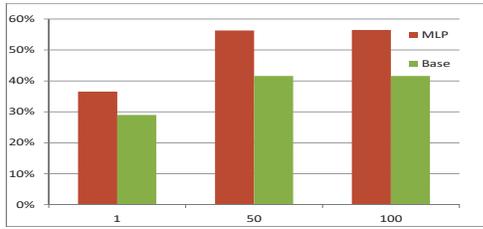

**Figure 8: Accuracy at Different Miles**

**Case Study** To illustrate the correctness of our model, we continue our examples. Specifically, we show the location assignments for some following relationships of user 13069282 in Tab. 5. Due to the space limitation, we remove the state information for each city in the table. Our method correctly assigns different locations (*e.g.*, Austin or Los Angeles) to her following relationships. Based on these assignments, *MLP* can estimate the user's multiple locations, *i.e.*, Los Angeles and Austin, correctly. In addition, it allows us to group a user's followers into different geo groups (*e.g.*, Los Angeles and Austin). Geo groups can be further used to group followers into more meaningful groups (*e.g.*, classmates in Austin).

**Table 5: Case Studies on Relationship Explanation**

| User ID: 13069282, Location: Los Angeles | | | |
|---|---|---|---|
| Follower's ID and Follower's Location | | Location Assignments | |
| ID | Location | User | Follower |
| 101566144 | Austin | Austin | Austin |
| 14119630 | Portland | Los Angeles | Los Angeles |
| 15669188 | Los Angeles | Los Angeles | Los Angeles |
| 53154473 | Long Beach | Los Angeles | Long Beach |

## 6. CONCLUSION

In this paper, we propose *MLP* to profile locations for Twitter users and their relationships with their following network and tweets. To the best of our knowledge, *MLP* is the the first model that 1) discovers users' multiple locations and 2) profiles both users and relationships. Specifically, for profiling users' locations, *MLP* advances the existing methods from the following aspects: 1) it profiles a user's home location more accurately, as it fundamentally models that following relationships and tweeted venues are generated by users' multiple locations, and may be even noisy, and 2) it profiles a user's locations more completely, as it explicitly models that a user has multiple locations. In addition, *MLP* is able to profile each following relationship in terms of users' hidden locations, and reveals the true geo connection in the relationship. We also conduct extensive experiments on a large-scale data set and demonstrate those advantages.